%% file: main.tex
\pdfoutput=1

\documentclass[conference]{IEEEtran}
\IEEEoverridecommandlockouts
\usepackage{cite}
\usepackage{amsmath,amssymb,amsfonts}
\usepackage{algorithmic}
\usepackage{graphicx}
\usepackage{textcomp}
\usepackage{xcolor}
\usepackage{caption,subcaption}
\usepackage{multirow}

\usepackage{hyperref}

\def\BibTeX{{\rm B\kern-.05em{\sc i\kern-.025em b}\kern-.08em
    T\kern-.1667em\lower.7ex\hbox{E}\kern-.125emX}}

\newcommand\blfootnote[1]{%
  \begingroup
  \renewcommand\thefootnote{}\footnote{#1}%
  \addtocounter{footnote}{-1}%
  \endgroup
}
    
\begin{document}

\title{Exploring the data of blockchain-based metaverses}

\author{
\IEEEauthorblockN{S. Casale Brunet$^{1,2}$, M. Mattavelli$^{1}$, L. Chiariglione$^{3}$}
\IEEEauthorblockA{École Polytechnique Fédérale de Lausanne, Switzerland\\
WhaleAnalytica.com, Switzerland\\
CEDEO SRL, Italy}
}

\maketitle

\input{tex/abstract}
\blfootnote{\scriptsize\copyright 2023 IEEE - Personal use of this material is permitted. Permission from IEEE must be obtained for all other uses, in any current or future media, including reprinting/republishing this material for advertising or promotional purposes, creating new collective works, for resale or redistribution to servers or lists, or reuse of any copyrighted component of this work in other works.}

\input{tex/introduction}
\input{tex/platforms}

\input{tex/data_aggregation}
\input{tex/data_analysis}

\input{tex/data_viz}

\input{tex/conclusions}

\bibliographystyle{IEEEtran}
%\bibliography{biblio}

\end{document}

%% file: tex/abstract.tex
\begin{abstract}
In recent years the concept of metaverse has evolved in the attempt of  defining richer immersive and interactive environments supporting various types of virtual experiences and interactions among users. This has led to the emergence of various different metaverse platforms that utilize blockchain technology and non-fungible tokens (NFTs) to establish ownership of metaverse elements and attach features and information to it. This article will delve into the heterogeneity of the data involved in these metaverse platforms, as well as highlight some dynamics and features of them. Moreover, the paper introduces a metaverse analysis tool developed by the authors, which leverages machine learning techniques to collect and analyze daily data, including blockchain transactions, platform-specific metadata, and social media trends. Experimental results are reported are presented with a use-case scenario focused on the trading of digital parcels, commonly referred to as metaverse real estate.
\end{abstract}


%% file: tex/introduction.tex
\section{Introduction}
The metaverse can be defined as a digital platform consisting of virtual environments and virtual worlds enabled by various technologies. These virtual environments can be created and customized by individuals or organizations for a variety of purposes, including entertainment, education, communication, and business. The metaverse can consist of multiple virtual layers, which can be connected through metachains and secured through the use of blockchain technology~\cite{metaverse_layers_2022,metachain_2022,meta_blockchain_definition_2022}. Its implementation may also require the use of technologies such as virtual reality, augmented reality, and artificial intelligence, depending on the specific use case~\cite{metaverse_ucs_2022}, with the human experience and interaction remaining a key component~\cite{metaverse_survey_2021}.

This paper focuses on a specific type of metaverse platform based on the concept of virtual land parcels, in which blockchain technology is used to enable ownership and representation of these digital assets. In other words, these platforms implement the concept of real estate in a distributed, trustless, and interactive environment.  A comprehensive understanding of this new digital asset class requires knowledge of topics such as traditional real estate and financial markets, blockchain technology, cryptocurrency assets, and non-fungible tokens (NFTs). Studies on blockchain-based metaverse platforms, such as Decentraland and The Sandbox Game, have shown that the location of virtual parcels is a key factor in determining their value, and that the market for digital land represented as NFTs is similar to the market for physical real estate~\cite{land_evaluation_2021,sandbox_report_2022, sandbox_report_v1_2022}.

This paper presents a technical analysis of the key components of blockchain-based metaverse platforms based on virtual land parcels. It illustrates and demonstrates how various data points, such as blockchain transactions, the number of users connected to the platform, and social media engagement, can be collected and effectively used to create accurate statistical models for determining the value of each individual parcel within each metaverse. In contrast to the state of the art, where studies focus on a specific platform and generally only consider transactions on the blockchain, this study presents a cross-sectional analysis of the top five Ethereum-based metaverses in which all collected heterogeneous data is analyzed on a daily basis, giving users the ability to assess the economic value of the parcels in these platforms.

The paper is structured as follows: Section~\ref{s:platforms} provides an overview of the main technical components of blockchain-based metaverse platforms based on virtual land parcels and NFTs; Section~\ref{s:processing} illustrates the different types of data that can be extracted and collected from these platforms; Section~\ref{s:analysis} analyzes the collected data and demonstrates how it can be used to build effective statistical models based on machine learning techniques to estimate the fair economic value of each parcel; Finally, Section~\ref{s:conclusions} concludes the paper and discusses future research directions.

%% file: tex/platforms.tex
\section{Blockchain-based metaverse enviroments}
\label{s:platforms}
Digital real estate markets within metaverses, also known as virtual worlds, often exhibit characteristics similar to traditional real estate markets, such as limited availability of land and the inability to easily move or transfer ownership of property~\cite{land_evaluation_2021}. However, these markets utilize decentralized technologies, such as blockchain and smart contracts, to facilitate secure and trustless transactions. This means that individuals can directly participate in the economy and own digital real estate, referred to as digital parcels, without the need for a central authority to verify or mediate the transaction. 

In this article, we examined five Ethereum-based platforms based on their popularity, trading volume and our own expertise. These are: Voxels, Decentraland, The Sandbox Game, Somnium Space and Otherside. It is important to point out that the contents of this list are derived entirely from the knowledge and expertise of the authors. It is important to note that this list should not be interpreted as providing any form of financial advice, as it is intended for informational purposes only.

%% file: tex/data_aggregation.tex
\section{Data collection}
\label{s:processing}
\begin{figure}[htbp]
\centering
\includegraphics[width=0.85\linewidth]{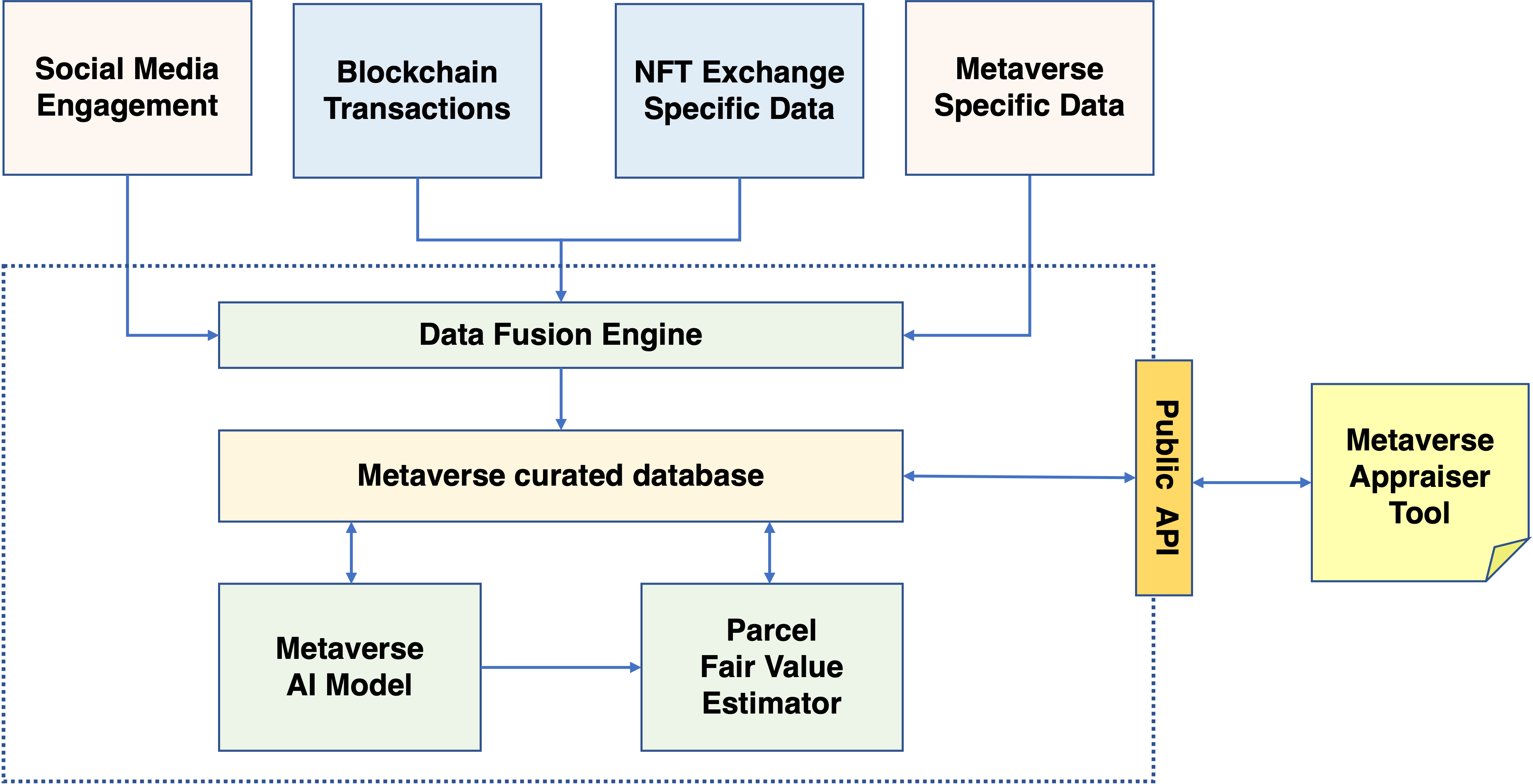}
\caption{Metaverse analysis frameworks developed in the paper.}
\label{f:pipeline}
\end{figure}

To fully understand and evaluate the value of virtual worlds in metaverse platforms, it is important to consider the types of data that can be analyzed for each environment. These data can be classified into two categories: on-chain data, which refers to financial transactions involving NFTs of parcels and are stored on the blockchain, and off-chain data such as parcel descriptions (e.g., location, size) and utilization (e.g., traffic patterns) that are generally not persistent and available from centralized servers. These data must be aggregated and carefully organized in order to be analyzed effectively. In the following sections, we will explore the main types of data that make up these metaverses and how we have implemented these data in our daily data acquisition and analysis tool, shown in Figure~\ref{f:pipeline}. We have made this tool publicly accessible~\cite{wa_metaverse_url} and we have also developed an API that allows users to retrieve heterogeneous data from various metaverses with a common semantics, ensuring that the data is always up-to-date. In the following, we present a discussion on the various types of data.
a) \textit{Metaverse-specific data}: information about each parcel in a virtual world, including its location and maximum build size, is often stored on centralized servers and represented as JSON files. This information, known as metadata, is usually encoded according to the \textit{de-facto} ERC721 metadata standard or the Enjin metadata recommendations. More advanced information about a parcel can typically be obtained through the metaverse platform's public API (e.g., see~\cite{decentralandAPI} for Decentraland, and~\cite{cryptovoxelsAPI} for Voxels).
b) \textit{Blockchain transactions}: the data stored on public blockchains, which are publicly accessible by design, can be efficiently accessed when structured in formats such as SQL~\cite{ethBrinckman}. The metaverse environments in this study are primarily based on Ethereum (with some secondary use of Polygon). The data collection techniques used are the ones we described in~\cite{scb_2021_erc721}. 
c) \textit{NFT exchange-specific data}: parcels can be traded on exchanges specific to the platform, such as the Decentraland Marketplace, and on more broadly deployed exchanges like OpenSea, LooksRare, and others. The information about the parcels that are on sale, including their price, is not stored on the blockchain but rather on the exchange website. To keep track of this information, it may be necessary to use the exchange API (if available, e.g., OpenSea API~\cite{opensea_api}) or web scraping techniques. Two interesting insights to consider when looking at this list of parcels for sale are the lowest price on the list, also known as the floor price, and the size of the list. The size of the list, along with the number of daily transactions, can give an indication of how 'liquid' the collection trading is. 
d) \textit{Media and social media popularity}
The popularity and social community of cryptocurrencies and NFT assets in mainstream and social media is a very important factor. In fact, studies such as~\cite{scb_2022_twitter} have emphasized this phenomenon. It is therefore important to monitor the sentiment on main social media platforms (e.g., Twitter, Reddit, Google). This can provide insight into the popularity of each metaverse platform and the broader concept of metaverse which, as we will see in the next sections, are correlated on the average price of the parcels.

%% file: tex/data_analysis.tex
\section{Data Analysis and Visualisation}
\label{s:analysis}
In the following, we describe how we analyzed the market for the five metaverses described in Section~\ref{s:platforms} for the period from January 1, 2021 to November 30, 2022. First, we will describe the techniques implemented for data collection, followed by the types of analysis carried out: starting with a global market analysis, then for each separate platform, and ending with the implementation of a machine learning model where, using the available data, it has been possible to define a suitable value for each land in the various metaverses.

\subsection{Dataset}
We obtained information on the blockchain transactions of the parcel NFTs and the average daily price of cryptocurrencies related to each metaverse (e.g., SAND and MANA) using the Dune platform~\cite{dune}. This platform provides SQL-structured and querable data on major blockchains, including raw transactions data and aggregated views (e.g., NFT trades). Using the official Twitter API, we collected data on social trends by gathering all tweets mentioning the accounts of each project and those containing the "\#metaverse" term hashtag, as well as the Google trend for the "metaverse" term. For each metaverse platform, specific information was then gathered based on the information available from their metadata. This is summarized in Table~\ref{t:data:features}. All of the resulting data and metadata we obtained were saved in our local database, as illustrated in Figure~\ref{f:pipeline} where the metaverse analysis framework we developed is shown.
Table~\ref{t:dataset} summarizes the volumes of trades in USD and the number of tweets. For the purpose of this study, we considered only transactions with an economic value (i.e., not those where only the owner of the token associated with the parcel has changed without a corresponding economic exchange). We also filtered these transactions by eliminating, for each project, those above the 99th percentile. According to the table, the total volume of transactions that we considered was approximately USD 1,500M and included approximately 160k transactions (with 10\% of the total volume and 1\% of the transactions already subtracted). At this stage, we did not perform any filtering on the collected tweets.
\input{tables/data.tex}
\input{tables/dataset.tex}

\subsection{Metaverse market trends}
During the period we studied, several notable events occurred: Facebook rebranded itself to Meta in late October 2021, leading to a surge in mainstream interest in the term "metaverse"; the rapid growth of the cryptocurrency market, driven primarily by Ethereum and Bitcoin~\cite{fintech1030017}, reached all-time highs in November 2021; the Otherside platform was launched on 30th April, 2022; successively the market as a whole saw contraction and crash in both equities and cryptocurrency due to challenging macroeconomic conditions. These events likely had an impact on the trend in digital land sales for the five metaverses we analyzed, as shown in Figure~\ref{f:trading:all}. %

\subsection{Platform-specific market trends}
\begin{figure}[htbp]
\captionsetup[subfigure]{aboveskip=-3pt,belowskip=3pt}
\centering

 \begin{subfigure}[b]{0.48\textwidth}
     \centering
     \includegraphics[width=\textwidth]{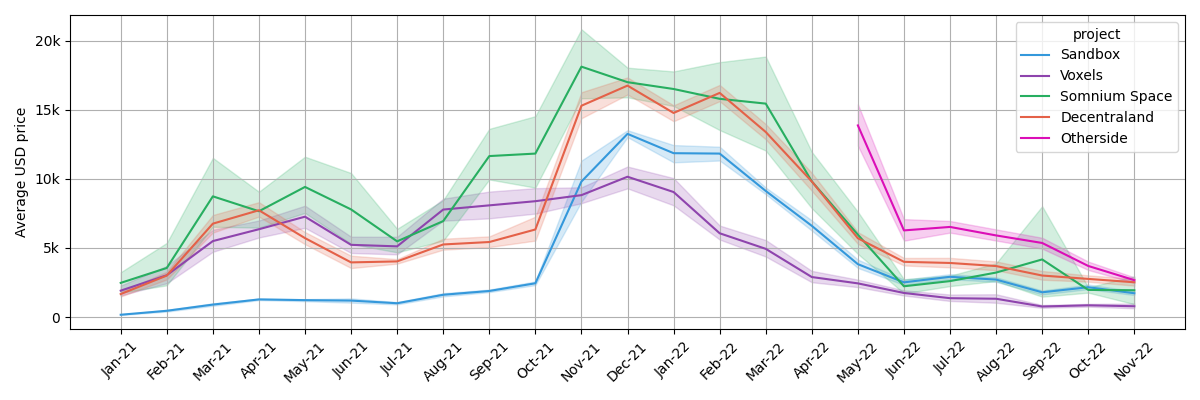}
     \caption{Parcels daily trades average price}
     \label{f:trading:avgprice}
 \end{subfigure}

 \begin{subfigure}[b]{0.48\textwidth}
     \centering
     \includegraphics[width=\textwidth]{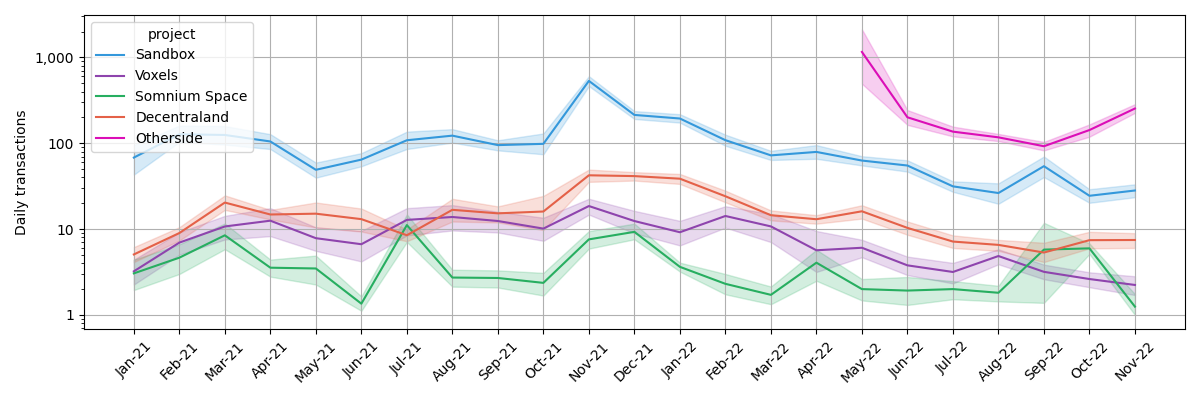}
     \caption{Number of parcel daily trades}
     \label{f:trading:trx}
 \end{subfigure}
 
 \begin{subfigure}[b]{0.48\textwidth}
     \centering
     \includegraphics[width=\textwidth]{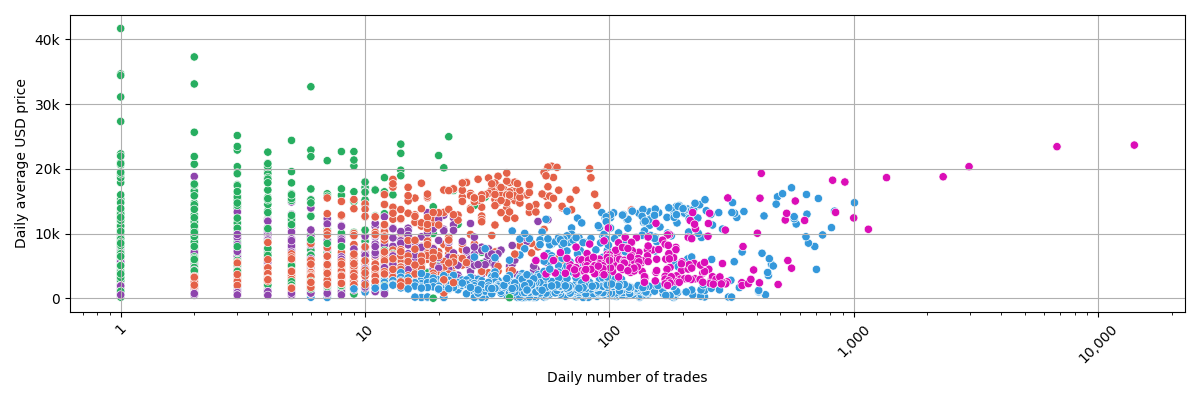}
     \caption{Parcels number of daily trades and average price distribution}
     \label{f:trading:trx_vs_price}
 \end{subfigure}

 \begin{subfigure}[b]{0.48\textwidth}
     \centering
     \includegraphics[width=\textwidth]{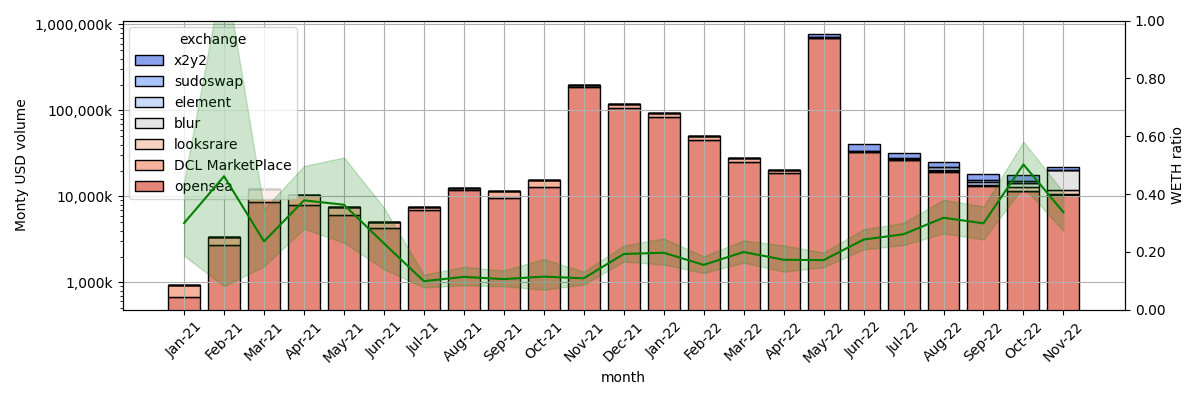}
     \caption{Parcels monthly volume per exchange. The green line, right y-axis, represents the ratio between WETH and other cryptocurrencies.}
     \label{f:trading:exchange}
 \end{subfigure}
 
\caption{Parcel trading volume and number of transactions.}
\label{f:trading:all}
 
\end{figure}

We can further delve into which metaverse platform had the most success in terms of trade volume and social media engagement by examining Figure~\ref{f:trading:all}. We can see that all collections saw the number of transactions and their average value increase following the explosion of interest in the metaverse topic in November 2021, and then followed the downward trend that began in spring 2022 (Figures~\ref{f:trading:trx} and~\ref{f:trading:avgprice}). The overall market considering all the five projects might not have been negatively impacted, however this is only due to the launch of "Otherside" by the creators of the BAYCs (which is one of the most successful and influential collection in the NFT market today). In fact, "Otherside" has managed to become one of the metaverse projects with the most traded volume in a short period of time (see Table~\ref{t:dataset}). It is interesting to see the distribution of daily transactions versus average daily price illustrated in Figure~\ref{f:trading:trx_vs_price}: from here, we can see that the market is clustered into two main groups, with "The Sandbox Game" and "Otherside" forming one group and the remaining collections forming the other. By analyzing the exchanges where these transactions take place, we estimated that approximately 88\% of the USD transaction volume occurs on OpenSea, while the next two most-used exchanges are x2y2 and the Decentraland (DCL) marketplace (note that in this latter only Decentraland parcels can be traded) with approximately 6\% and 3.6\%, respectively. We also find that ETH and WETH are the most common crypto-currencies used for trading, accounting for 80\% and 10\% of the total USD volume, respectively. WETH, which are ERC-20 tokens that represent 1:1 ETH, are often used to purchase parcels (and other NFTs) through a bidding process. Bids are usually placed below the current lowest price of the collection, known as the floor price. Once a parcel has been acquired, it may be resold in an attempt to make a (quick) profit. This is known as flipping. During times when the market is experiencing a negative trend, such as a liquidation phase, there may be an increase in the number of accepted bids for WETH. This can be seen in Figure~\ref{f:trading:exchange}, which shows the ratio (represented by the green line) between the daily trading volume of WETH and other currencies. This ratio tends to increase significantly when the market is experiencing a negative trend and average parcel prices are declining.

\subsection{Parcels position, geometry and traffic}
\begin{figure}[htbp]
\centering
\includegraphics[width=0.98\linewidth]{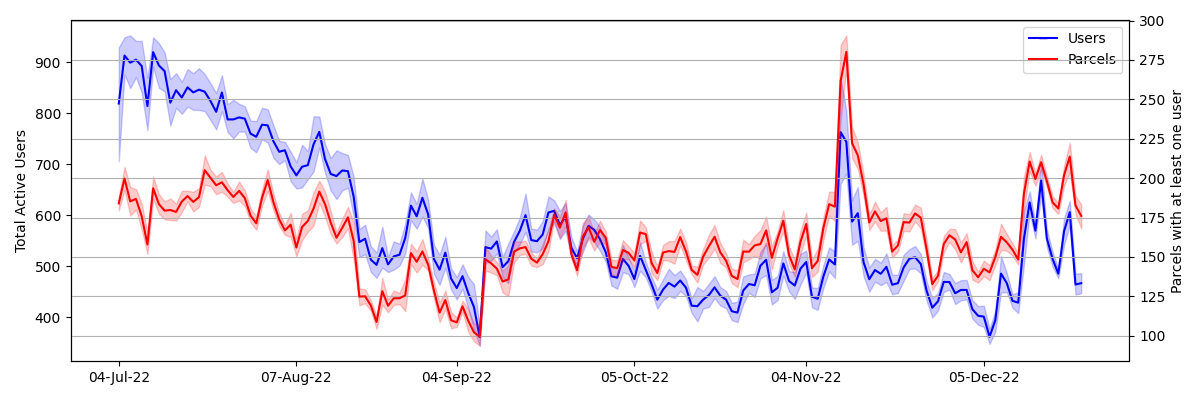}
\caption{Decentraland traffic traffic defined as maximum users connected per parcel per hour.}
\label{f:traffic}
\end{figure}

In the previous section, we analyzed various metaverses individually, examining the average daily price of parcels sold. If we instead focus on individual parcels, recent studies have shown that location is a key factor that can significantly impact the parcel value when compared to the average value. For example, studies~\cite{land_evaluation_2021} and~\cite{sandbox_report_v1_2022} on Decentraland and The Sandbox Game respectively have both concluded that, despite the absence of travel distance in the metaverse, location is extremely important. These studies, however, focus on two specific platforms where the size of each parcel is uniform. In the more general case of Voxels and Somnium Space parcel size may also affect the price of a parcel. Therefore, the framework we implemented (shown in Figure~\ref{f:pipeline}) also gathers the metadata for each parcel, including information about the available area and volume for construction on the parcel. In addition, for Decentraland, Somnium Space, and Voxels, we have also collected information about the traffic on each parcel. In the following, we analyze the information we have collected for each individual parcel in addition to their geographical location, as shown in Table~\ref{t:data:features}. 
a) \textit{Voxels}: each parcel has different dimensions, with associated height and area limits for building. 
For each parcel, we are able to obtain the daily cumulative number of unique users who have logged in.
b) \textit{Decentraland}: all the parcels have the same size of 16m x 16m, but adjacent parcels can be grouped into estates. As of now, there are approximately 2,160 estates. 
For each parcel, we are able to collect the number of concurrent users connected per hour: Figure~\ref{f:traffic} shows the maximum number of users connected to the platform from June 2022 to the end of November 2022 (the period for which we have data).
c) \textit{The Sandbox Game}: all the parcels have the same size of 96m x 96m, but adjacent parcels can be grouped into estates in fixed sizes of 3x3, 6x6, 12x12, and 24x24 parcels. 
d) \textit{Somnium Space}: there are three types of parcels with different sizes: 'S' (2,000m$^3$), 'M' (15,000m$^3$), and 'XL' (75,000m$^3$). 
For each plot, we collect the number of connected users per hour, distinguishing between spectators and players.
d) \textit{Otherside}: for each parcel, we identify sediments, artifacts, and the possible presence of one of the 10,000 Koda NFTs. 

\subsection{Machine learning models of metaverses}
\begin{figure}[htbp]
\centering
\captionsetup[subfigure]{aboveskip=-1pt,belowskip=3pt}
 \begin{subfigure}[b]{0.2\textwidth}
     \centering
     \includegraphics[width=\textwidth]{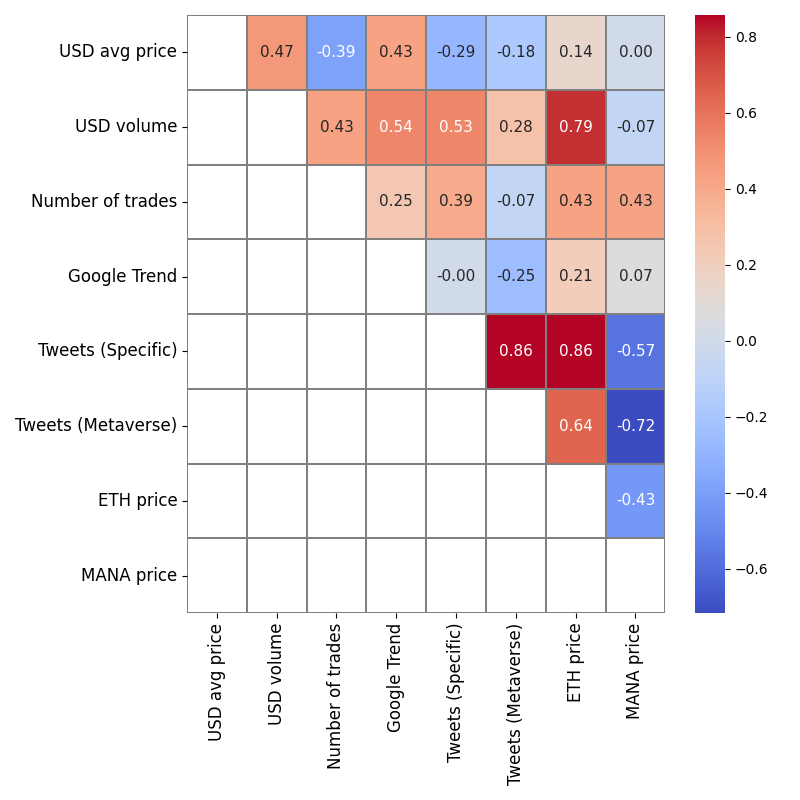}
     \caption{Decentraland}
     \label{f:correlation:decentraland}
 \end{subfigure}
 \begin{subfigure}[b]{0.2\textwidth}
     \centering
     \includegraphics[width=\textwidth]{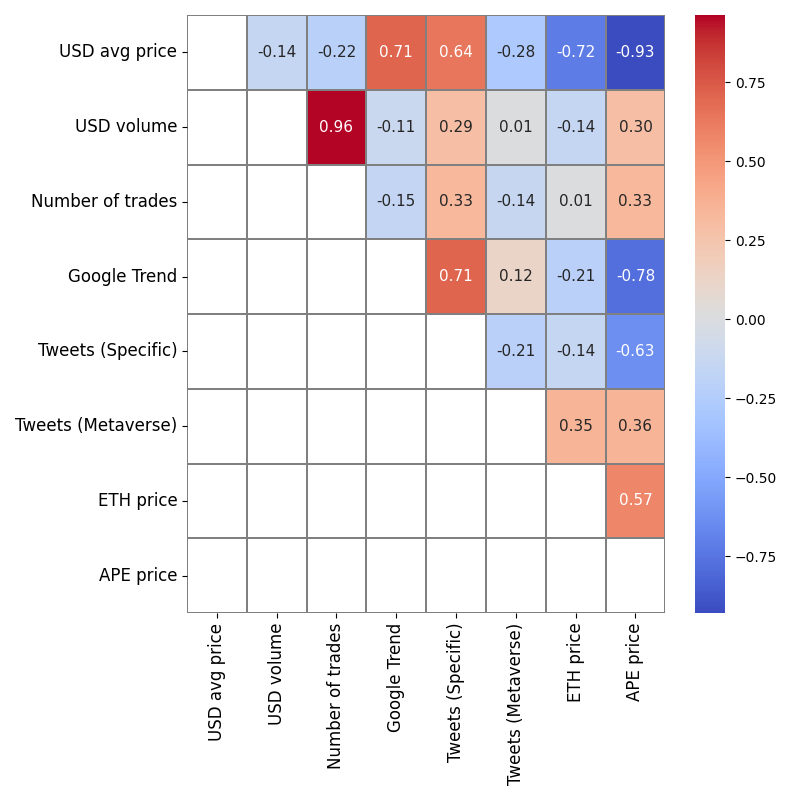}
     \caption{Otherside}
     \label{f:correlation:otherside}
 \end{subfigure}

\caption{Spearman correlations for some of the collected data.}
 \label{f:correlations}
\end{figure}

\begin{figure}[htbp]
\centering
\captionsetup[subfigure]{aboveskip=-2pt,belowskip=3pt}
 \begin{subfigure}[b]{0.4\textwidth}
     \centering
     \includegraphics[width=\textwidth]{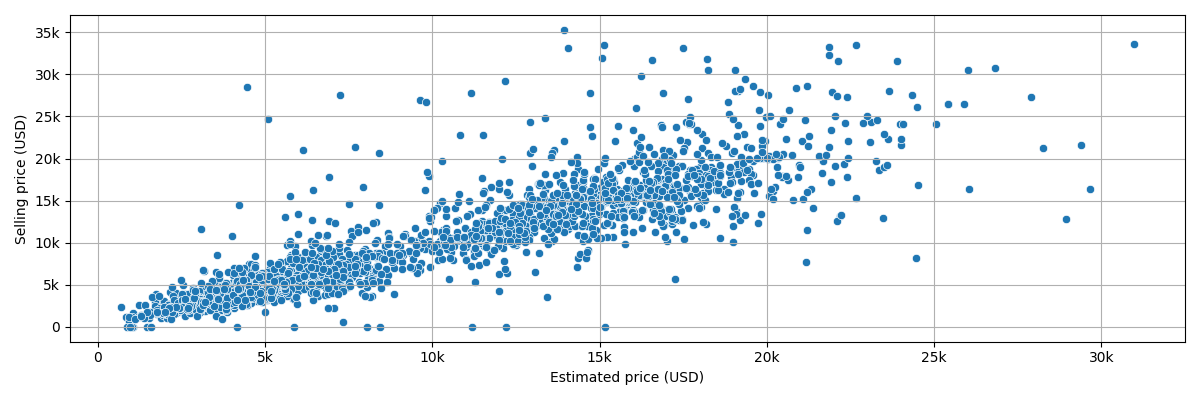}
     \caption{Estimated price (x-axis) v.s. actual sale price (y-axis)}
     \label{f:ai:decentraland:price}
 \end{subfigure}

  \begin{subfigure}[b]{0.4\textwidth}
     \centering
     \includegraphics[width=\textwidth]{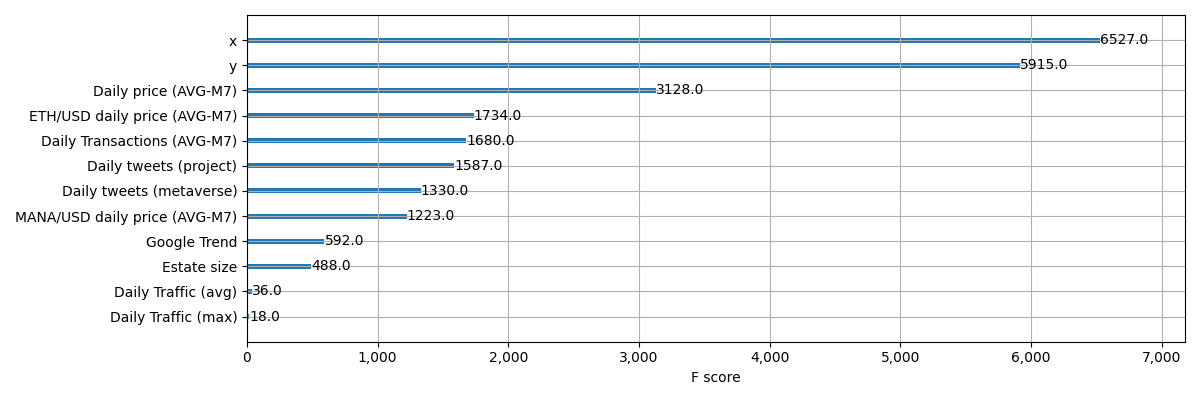}
     \caption{Features importance}
     \label{f:ai:decentraland:features}
 \end{subfigure}

\caption{Decentraland parcel price estimation model.}
\label{f:ai:decentraland}
 
\end{figure}

To examine potential correlations and build a statistical model to determine the economic value of each parcel, we first collected and organized data from various sources and at different levels. We then used a Spearman correlation analysis to analyze the following variables: daily average price, volume, and number of parcel sales; metaverse topic popularity on Google (measured through Google Trends); daily tweets related to the specific metaverse platform and the metaverse topic in general; and the daily dollar price of ETH and any platform's cryptocurrency. The results are displayed in Figure~\ref{f:correlations}. In order to improve the accuracy of the analysis, we first removed any seasonal components from each time series. 
We can see that the average price and trading volume are strongly correlated with the number of tweets for the Otherside platform (see Figure~\ref{f:correlation:otherside}), while for the other projects, there seems to be a stronger link with popularity on other channels, as indicated by the correlation with Google Trends (e.g., see Figure~\ref{f:correlation:decentraland}). This probably indicates that Twitter is less influential for NFT metaverse projects compared to what was observed for example in~\cite{scb_2022_twitter} for NFT profile pictures (PFP) projects. We believe that the current nature of Otherside's trading and its underdeveloped gaming environment make it more akin to a PFP project rather than a metaverse one. 
The second step was to understand in more detail which variables most influence the selling price of a plot. To do this, we used XGBoost (eXtreme Gradient Boosting)~\cite{XGBoost}, a widely-used machine learning algorithm that is particularly effective for regression and classification tasks. 
We conducted separate experiments for each platform, training the model to predict the prices of the plots based on the other available data. We randomly divided the dataset described in Table~\ref{t:dataset} into two parts: a training set containing 80\% of the transactions for each platform, and a test set containing the remaining 20\%. We then evaluated the model's accuracy and reliability using the test set by comparing its predictions to the actual sale prices of each plot transaction (e.g., see Figure~\ref{f:ai:decentraland:price}). A randomized search for hyperparameter tuning was used to identify the best parameters configuration for each model. The number of features and accuracy of each metaverse model are summarized in Table~\ref{t:ai:models}. In general, we found that parcel location (in terms of x, y coordinates) is the factor that most influences the sale price on each metaverse, as already demonstrated in~\cite{land_evaluation_2021,sandbox_report_2022, sandbox_report_v1_2022}. However, we can add that other factors with a significant influence on the selling price of a parcel are the average daily price of other plots sold, the daily price of ETH (which can also serve as a general crypto market indicator), and the level of activity on a parcel, as for example in the case of Decentraland (see Figure~\ref{f:ai:decentraland:features}). The results of this study indicate that user traffic on a parcel is not a significant determinant of its price. Instead, factors related to the revenue-generating potential of the parcel are more likely to play a role. In our opinion, this is because we are currently in an exploratory phase of the market, where individuals and organizations investing in digital parcels are primarily focused on acquiring strategic locations as a form of marketing investment.

\input{tables/ai.tex}

%% file: tables/data.tex
\begin{table}[]
\caption{Collected and analysed features for each metaverse platform.}
\label{t:data:features}
\centering
\resizebox{0.48\textwidth}{!}{
\begin{tabular}{|r|lllll|}
\hline
\multicolumn{1}{|l|}{} & \multicolumn{1}{l|}{\textbf{The Sandbox Game}} & \multicolumn{1}{l|}{\textbf{Decentraland}} & \multicolumn{1}{l|}{\textbf{Voxels}} & \multicolumn{1}{l|}{\textbf{Somnium Space}} & \textbf{Otherside} \\ \hline
\textbf{Blockchain transactions} & \multicolumn{5}{l|}{\begin{tabular}[c]{@{}l@{}}- Daily, weekly, monthly volume in ETH, USD, and metaverse specific currencies\\ - Daily, weekly, monthly number of transactions\\ - Daily, weekly, monthly parcel average price\\ - Daily ETH/USD and metaverse specific currencies  spot price\end{tabular}} \\ \hline
\textbf{Exchange information} & \multicolumn{5}{l|}{\begin{tabular}[c]{@{}l@{}}- Daily listed parcels\\ - Daily listings price distribution\end{tabular}} \\ \hline
\textbf{Parcel metadata} & \multicolumn{1}{l|}{\begin{tabular}[c]{@{}l@{}}- X, Y coordinates\\ - Proximity to POIs\\ - Estate size\end{tabular}} & \multicolumn{1}{l|}{\begin{tabular}[c]{@{}l@{}}- X, Y coordinates\\ - Proximity to POIs\\ - Estate size\end{tabular}} & \multicolumn{1}{l|}{\begin{tabular}[c]{@{}l@{}}- X, Y coordinates\\ - Proximity to POIs\\ - Volume  \end{tabular}} & \multicolumn{1}{l|}{\begin{tabular}[c]{@{}l@{}}- X, Y coordinates\\ - Area  \end{tabular}} & \begin{tabular}[c]{@{}l@{}}- X, Y coordinates\\ - Sediment and Artifact\\ - Koda\end{tabular} \\ \hline
\textbf{Parcel traffic} & \multicolumn{1}{l|}{} & \multicolumn{1}{l|}{- Hourly max. users} & \multicolumn{1}{l|}{- Daily cumulative users} & \multicolumn{1}{l|}{- Hourly max. users} &  \\ \hline
\textbf{Social media popularity} & \multicolumn{5}{l|}{\begin{tabular}[c]{@{}l@{}}- Number of daily tweets for topic and discussion related to the specific platform and broader metaverse topic \\ - Google popularity related to the metaverse topic\end{tabular}} \\ \hline
\end{tabular}
}
\end{table}

%% file: tables/dataset.tex
\begin{table}[]
\caption{Dataset size considering the time period from 1st January 2021 to 30th November 2022 included. Statistics on discarded data refer to the value of the project.}
\label{t:dataset}
\resizebox{0.48\textwidth}{!}{
\begin{tabular}{|r|rr|rr|rr|rr|rr|}
\hline
\multirow{2}{*}{\textbf{Metavers}} & \multicolumn{4}{c|}{\textbf{Volume (USD)}} & \multicolumn{4}{c|}{\textbf{Transactions}} & \multicolumn{2}{c|}{\multirow{2}{*}{\textbf{Tweets}}} \\ \cline{2-9}
 & \multicolumn{2}{c|}{\textit{\textbf{Considered}}} & \multicolumn{2}{c|}{\textit{\textbf{Discarded}}} & \multicolumn{2}{c|}{\textit{\textbf{Considered}}} & \multicolumn{2}{c|}{\textit{\textbf{Discarded}}} & \multicolumn{2}{c|}{} \\ \hline

Otherside &906,704,668   & 59.14\% & 138,340,545  & 13.24\% & 67,563   & 42.12\% &683  & 1.00\% & 1,015,177    & 1.73\%  \\ \hline
Voxels &33,393,758   & 2.18\% & 3,378,292  & 9.19\% & 5,565   & 3.47\% &58  & 1.03\% & 493,085    & 0.84\%  \\ \hline
Decentraland &112,433,708   & 7.33\% & 8,290,684  & 6.87\% & 11,218   & 6.99\% &114  & 1.01\% & 4,151,326    & 7.07\%  \\ \hline
Somnium Space &22,311,548   & 1.46\% & 2,231,310  & 9.09\% & 2,210   & 1.38\% &23  & 1.03\% & 129,348    & 0.22\%  \\ \hline
The Sandbox Game &458,337,419   & 29.89\% & 27,184,674  & 5.60\% & 73,844   & 46.04\% &746  & 1.00\% & 5,046,455    & 8.59\%  \\ \hline
Total &1,533,181,103 & & 179,425,508 & 10.48\% & 160,400 & & 1,624 & 1.00\% & 58,740,001 &  \\ \hline

 \end{tabular}
}
\end{table}

%% file: tables/ai.tex
\begin{table}[]
\caption{Parcel price estimation model size and accuracy.}
\label{t:ai:models}
\centering
\resizebox{0.40\textwidth}{!}{
\begin{tabular}{|r|c|cc|}
\hline
\multicolumn{1}{|r|}{\multirow{2}{*}{\textbf{Metaverse}}} & \textbf{Model Size} & \multicolumn{2}{c|}{\textbf{Accuracy \%}} \\ \cline{2-4} 
\multicolumn{1}{|c|}{} & \textit{\textbf{Number of features}} & \multicolumn{1}{c|}{\textit{\textbf{Training}}} & \textit{\textbf{Performance}} \\ \hline
\textbf{The Sandbox Game} & 8 & \multicolumn{1}{c|}{95.8} & 86.0 \\ \hline
\textbf{Decentraland} & 11 & \multicolumn{1}{c|}{91.3} & 88.7 \\ \hline
\textbf{Voxels} & 21 & \multicolumn{1}{c|}{98.0} & 75.1 \\ \hline
\textbf{Somnium Space} & 12 & \multicolumn{1}{c|}{95.1} & 88.3 \\ \hline
\textbf{Otherside} & 11 & \multicolumn{1}{c|}{87.5} & 71.5 \\ \hline
\end{tabular}
}
\end{table}

%% file: tex/data_viz.tex
\subsection{Data exploration and visualization tool}
\label{s:analysis:dataviz}
We have created an exploration tool to help users navigate and explore data more easily. This tool is available at~\cite{wa_metaverse_url} and allows users to browse different platforms at various levels, displaying various types of information directly on the plots. These include:
1) Land view, which colors parcels based on their characteristics in the metaverse (e.g., size); 2) Trading view, which highlights parcels for sale on different exchanges with different colors based on their sale price; 3) Last price view, which colors parcels based on the last sale price; 4) Value view, which colors parcels based on the ratio between the sale price and our estimated price; 5) Fair value view, which colors parcels based on our estimated fair value price; 6) Flip view, which uses color to indicate how many times the parcel has been traded. Depending on the structure of a particular metaverse, there may also be specific metrics such as: 7) Traffic view, which uses color to highlight the most heavily trafficked parcels; and 8) Resources view, which uses color to indicate the availability of different resources.

%% file: tex/conclusions.tex
\section{Conclusions}
\label{s:conclusions}
In this article, we conducted a technical analysis of five major blockchain-based metaverse platforms that implement the concept of digital land and real estate in a decentralized digital environment. We described the various technological components and how various types of data, such as blockchain transactions, parcel traffic, and engagement on social networks, can be effectively extracted and analyzed from these platforms. 
The results obtained have been: 1) the development of the first cross-platform metaverse data collection and analysis tool, data that is now accessible through a public and unified API; 2) the systematic creation of machine learning models that, through data fusion and curation, are able to estimate a fair value of each individual parcel in each metaverse; 3) the verification, thanks to these models, that location is a generally fundamental factor in determining the value of a parcel, as already demonstrated bysome state-of-the-art work. In comparison to these studies, which only focus on two specific platforms, our work has been performed on the five main Ethereum-based platforms.
Future studies will aim to improve the accuracy of these estimation models and study more complex traffic patterns, for example by testing whether it is possible to distinguish between real users and bots.